\documentclass[reqno]{amsart}
\usepackage{tabularx}
\usepackage{graphicx,graphics}
\usepackage{subfig}
\usepackage{url}

\newcommand{\Real}{\mathbb R}
\newcommand{\abs}[1]{\left\vert#1\right\vert}

\numberwithin{thm}{section}
\numberwithin{lem}{section}
\numberwithin{coll}{section}
\numberwithin{rem}{section}
\numberwithin{exm}{section}
\numberwithin{prop}{section}
\numberwithin{equation}{section}

\numberwithin{equation}{section}

\begin{document}
\centerline {\textsc {\large General Extreme Value Modeling and Application of Bootstrap }}
\centerline {\textsc{ on  Rainfall Data - A Case Study}} 
\vspace{0.5in}
\begin{center}
 Ali Saeb\footnote{Corresponding author: ali.saeb@gmail.com}\\
Theoretical Statistics and Mathematics Unit,\\ Indian Statistical Institute, Delhi Center,\\7 S.J.S Sansanwal Marg, New Delhi 110016, India
\end{center}

\vspace{1in}


\noindent {\bf Abstract:} 
Extreme value theory is concerned with probabilistic and statistical questions related to very high or
very low values in sequences of random variables and in stochastic processes. The subject has a rich
mathematical theory and also a long tradition of applications in a variety of areas. Among many
excellent books on the subject, Coles [2] while the book by concentrates on data analysis and
statistical inference for extremes. In this article, we present a case study wherein we model annual
maximum yearly rainfall data using the generalized extreme value distribution. 
 Also, we use R software for
data analysis and give the R codes in the appendix.

\vspace{0.5in}

\vspace{0.2in} \noindent {\bf Keywords:} General Extreme Value Distribution, bootstrap, Jackknife method, Return level, AIC.

\vspace{0.2in} \noindent {\bf MSC 2010 classification:} 60G70, 62G32.
\newpage
\section{Introduction}
The limit laws of linearly normalized partial maxima $M_n=X_1\vee\ldots\vee X_n$ of independent and identically distributed (iid) random variables (rvs) $X_1,X_2,\ldots,$ with common distribution function (df) $F,$ namely,
\begin{equation}\label{Introduction_e1}
	\lim_{n\to\infty}\Pr(M_n\leq a_nx+b_n)=\lim_{n\to\infty}F^n(a_nx+b_n)=G(x),\;\;x\in \mathcal{C}(G),
\end{equation}
	where, $a_n>0,$ $b_n\in\Real,$ are norming constants, $G$ is a non-degenerate df, $\mathcal{C}(G)$ is the set of all continuity points of $G,$ are called max stable laws. If, for some non-degenerate df $G,$ a df $F$ satisfies (\ref{Introduction_e1}) for some norming constants $a_n>0,$ $b_n\in\Real,$ then we say that $F$ belongs to the max domain of attraction of $G$ under linear normalization and denote it by $F\in \mathcal{D}(G).$ Limit dfs $G$ satisfying (\ref{Introduction_e1}) are the well known extreme value types of distributions, or max stable laws, namely,
	\begin{eqnarray*}
			\text{I. the Fr\'{e}chet law:} & \Phi_\alpha(x)  = \left\lbrace	
							\begin{array}{l l}
							 0, &\;\;\; x< \mu, \\
							 e^{-\left(\frac{x-\mu}{\sigma}\right)^{-\alpha}}, &\;\;\; \mu\leq x;\\
							 \end{array}
							 \right. \\					
		\text{II. the Weibull law:} & \Psi_\alpha(x) =  \left\lbrace
						\begin{array}{l l} e^{- |\frac{x-\mu}{\sigma}|^{\alpha}}, & x<\mu, \\
						1, & \mu\leq x;
						\end{array}\right. \\
		\text{III. the Gumbel law:} & \Lambda(x) = e^{-e^{-\left(\frac{x-\mu}{\sigma}\right)}};\;\;\;\;\; x\in\Real;
	\end{eqnarray*}
$\alpha,\sigma>0,$ $\mu\in\Real$ being a parameters. Criteria for $F\in \mathcal{D}(G)$ are well known (see, for example, Galambos, 1987; Resnick, 1987; Embrechts et al., 1997).

	The three types of distributions may all be represented as members of a
single family of generalized extreme value distributions with df 
\begin{equation}\label{gev}
	G(x)=e^{-\left(1+\xi\left(\frac{x-\mu}{\sigma}\right)\right)^{-1/\xi}},\;\;\Big\{x: 1+\xi\left(\frac{x-\mu}{\sigma}\right)>0\Big\},
\end{equation}
where, $\mu\in\Real,$ $\sigma>0$ and $\xi\in\Real.$ This is the generalized extreme value (GEV) family of distributions. The type I and type II classes of extreme value distributions correspond respectively to the case $\xi>0$ and $\xi<0$ in this parametrization. The subset of the GEV family with $\xi=0$ is interpreted as the limit of (\ref{gev}) as $\xi\to 0,$ leading to the Gumbel family with df
\[G(x)=e^{-e^{-\left(\frac{x-\mu}{\sigma}\right)}},\;\;x\in\Real.\]
Gilleland and Katz (2005), designed the Extremes Toolkit ("extRemes") to facilitate the use of extreme value theory in applications. Also, "ismev" is another R package which includes functions to support the computations carried out by Coles (2001). Castilo et al. (2004) and Coles (2001) are good references to the application of extreme value distribution.

In this article, we model the annual maximum yearly rainfall data of station Eudunda, Australia which collected during 1881-2013 and using the GEV family of dfs by motivate the methods are discussed in Coles (2001) by consider the stationary model. The method of bootstrap for estimate the bias is discussed Efron and Tibshirani (1993). We use the packages of  "ismev" and "bootstrap" installed in R software and give the R codes in the appendix. Last, we introduce the new function to find out the bias and standard error with Jackknife method for GEV distributions.

\section{Methodology}
\textbf{Principles of estimation.} It is desirable that estimates are close to the parameter value they are estimating. We assume a vector of parameters such as $\theta.$
The bias of $\hat{\theta}$ as an estimate of an estimator $\hat{\theta}_0$ of $\theta_0$ is defined by
\begin{eqnarray}\label{bias}
\texttt{bias}=E{\hat{\theta}_0}-\theta_0.
\end{eqnarray}
A large bias is usually an undesirable aspect of an estimator's performance. We can use the bootstrap to assess the bias of any estimator $\hat{\theta}_0.$ We generate $B$ independent bootstrap samples $X^{*1},X^{*2},\ldots,X^{*B},$ each consisting of $n$ data values drawn with replacement from $X,$ as $X^{*1}=X_{i_1},X^{*2}=X_{i_2},\ldots,X^{*n}=X_{i_n}.$ We can select the sample size of $B$ in the range $25-200,$ (see, Efron and Tibshirari (1986)). Then, evaluate the bootstrap replication corresponding to each bootstrap sample, it may be an indication that the statistic $\hat{\theta}^*(b)=S(X^{*b}),\;b=1,2,\ldots,B.$ The bootstrap estimate of bias is defined by
\[\text{bias}_B=\hat{\theta}_0^{*}-\hat{\theta}_0,\]
where, $\hat{\theta}_0^{*}=\frac{1}{n}\sum_{b=1}^{B}\hat{\theta}^{*}(b).$  Now, we have concentrated on standard error as a measure of accuracy for an estimator $\hat{\theta}.$ Estimate the standard error $\texttt{se}_B(\hat{\theta})$ by the sample standard deviation of the $B$ replications,
\begin{eqnarray}
\hat{\texttt{se}}_B=\left[\sum_{b=1}^{B}[\hat{\theta}^*(b)-\hat{\theta}_{0}^{*}]^2/(B-1)\right]^{\frac{1}{2}}.
\end{eqnarray}

The Jackknife estimate of bias is another method to find out the bias which it was original computer based method for estimating biases and standard errors. Suppose we have a sample $X=(x_1,\ldots, x_n)$ and an estimator $\hat{\theta}_0=S(X).$ The $i^{th}$ Jackknife sample $x_{(i)},$ is defined to be $x$ with the $i^{th}$ data point removed,
\[x_{(i)}=(x_1,x_2,\ldots, x_{(i-1)}, x_{(i+1)},\ldots, x_n),\]
for $i=1,2,\ldots,n.$ The Jackknife estimate of bias is defined by
\[\texttt{bias}_{jack}=(n-1)(\hat{\theta}_{(\cdot)}-\hat{\theta}_0),\]
where, $\hat{\theta}_{(\cdot)}=\sum_{i=1}^{n}\hat{\theta}_{(i)}/n.$
The Jackknife estimate of standard error is,
\[\hat{\texttt{se}}_{jack}=\left[\frac{n-1}{n}\sum_{i=1}^{n}\left(\hat{\theta}_{(i)}-\hat{\theta}_{(\cdot)}\right)^2\right]^{1/2}.\]
The jackknife often provides a simple and good approximation to the bootstrap, for estimation of standard errors and bias.

As a rule of thumb, a bias of less than $0.25$ standard errors can be ignored, unless we are trying to do careful confidence interval calculations. The root mean square error of an estimator $\hat{\theta}$ for $\theta,$ is $\sqrt{E(\hat{\theta}-\theta)^2},$ a measure of accuracy that takes into account both bias and standard error. It can be shown that the root mean square equals,
\begin{eqnarray}
\sqrt{E(\hat{\theta}_0-\theta_0)^2}&=&\hat{\texttt{se}}
\sqrt{1+\left(\frac{\text{bias}}{\hat{\texttt{se}}}\right)^2},\nonumber\\
&\simeq&\hat{\texttt{se}}
\left[1+0.5\left(\frac{\text{bias}}{\hat{\texttt{se}}}\right)^2\right].
\end{eqnarray}
If $\text{bias}=0$ then the root mean square equals its minimum value of standard error. If $\abs{\text{bias}/\texttt{se}}<0.25,$ then the root mean square error is no more that about $0.031$ greater than value of standard error.

The obvious bias corrected estimator is,
\[\theta_{corr}=\hat{\theta}_0-\texttt{bias}=2\hat{\theta}_0-\hat{\theta}^*(.),\] where, $\text{bias}=\text{bias}_B.$
Correcting the bias may cause a larger increase in the standard error, which in turn results in a larger root mean square error.
If bias is small compared to the estimated standard error $\hat{\texttt{se}},$ then it is safer to use $\hat{\theta}_0$ than $\theta_{\texttt{corr}}.$ If bias is large compared to standard error, then it may be an indication that the statistic $\hat{\theta}_0=S(X)$ is not an appropriate estimate of the parameter $\theta.$ For more details of these specific for bootstrap method, see, Efron and Tibshirani (1993) and Davison and Hinkley (1997).

Quantifying the precision of an estimator can usually be made more explicit by calculating a confidence interval. A standard result says that $\hat{\theta}_0$ is the maximum likelihood estimator has a limiting multivariate normal distribution with mean $\theta_0$ and variance covariance matrix $V_{\theta_0}=I(\theta_0)^{-1},$ where,
\[I(\theta) = \left[ \begin{array}{ccccc}
e_{1,1}(\theta) & &\cdots & &e_{1,d}(\theta) \\
&\ddots&e_{i,j}(\theta)&&\\
\vdots &  e_{j,i}(\theta)& & &\vdots \\
&&&\ddots&\\
e_{d,1}(\theta) &\cdots &&\cdots&e_{d,d}(\theta) \end{array} \right],\] 
 with $e_{i,j}(\theta)=-E\frac{\partial^2\ell(\theta)}{\partial \theta_i \partial\theta_j},$ and $\ell(\theta)=\sum_{i=1}^{n}\log f_\theta(x_i)$ is log likelihood function. The matrix $I(\theta)$ is "expected information matrix". Since the true value of $\theta_0$ is generally unknown, it is usual to approximate the term of $I$ with those of the "observed information matrix", defined by
\[I_O(\theta) = \left[ \begin{array}{ccccc}
-\frac{\partial^2\ell(\theta)}{\partial\theta_1^2} & &\cdots & &-\frac{\partial^2\ell(\theta)}{\partial \theta_1 \partial\theta_d} \\
&\ddots&-\frac{\partial^2\ell(\theta)}{\partial \theta_i \partial\theta_j}&\\
\vdots &  -\frac{\partial^2\ell(\theta)}{\partial \theta_j \partial\theta_i}& & &\vdots \\
&&&\ddots&\\
-\frac{\partial^2\ell(\theta)}{\partial \theta_d \partial\theta_1} &\cdots &&\cdots&-\frac{\partial^2\ell(\theta)}{\partial \theta_d^2} \end{array} \right],\] 
 
and evaluated at $\theta=\hat{\theta}.$ Denoting an arbitrary term in the inverse of $I_O(\theta)$ by $\tilde{\sigma}_{i,j},$ it follows that an approximate $(1-\tau)$ wehere, $0<\tau<1,$ confidence interval for $\theta_i$ is, 
\[\hat{\theta}_i\pm z_{\frac{\tau}{2}}\sqrt{\tilde{\sigma}_{i,i}}.\]

Let $\hat{\theta}_0$ be the maximum likelihood estimator of the $d-$ dimensional parameter $\theta_0$ with approximate variance covariance matrix $V_{\theta_0}.$ If $\eta=g(\theta)$ is a scalar function, the maximum likelihood estimator of $\eta$ is $\hat{\eta}=g(\hat{\theta}_0),$ then $\hat{\eta}\to_d N(g(\theta), V_{g(\theta)})$ where, $V_{g(\theta)}=\nabla \eta^T\,V_{\theta}\,\nabla\eta$ with $\nabla \eta=[\frac{\partial \eta}{\partial \theta_1},\ldots,\frac{\partial \eta}{\partial \theta_d}]$ evaluated at $\hat{\theta}.$ For more details see, Casella and Berger (2002).
 
Alternatively, a confidence interval can derived from the likelihood function, by using approximation
\[D(\theta_0)=2(\ell(\hat{\theta}_0)-\ell(\theta_0))\sim \chi_d^2,\]
It follows that an approximate $(1-\tau)$ confidence region for $\theta_0$ is given by
\[C_\tau=\{\theta: D(\theta)\leq c_\tau\},\]
where, $c_\tau$ is the $(1-\tau)$ quantile of the $\chi_d^2$ distribution. This approximation is usually more accurate than that based on the asymptotic normality of the maximum likelihood estimator.

The log likelihood for $\theta$ can be formally written as $\ell(\theta_i,\theta_{-i})$ where, $\theta_{-i}$ denotes all components of $\theta$ excluding $\theta_i.$ The profile log likelihood for $\theta_i$ is defined as
\[\ell_p(\theta_i)=\max \ell(\theta_i,\theta_{-i}).\] That is, for each value of $\theta_i,$ the profile log-likelihood is the maximized log likelihood with respect to all other component of $\theta.$ This definition generalizes to the situation where $\theta$ can be partitioned into two component, $(\theta^{(1)},\theta^{(2)}),$ of which $\theta^{(1)}$ is the $k-$ dimensional vector of interest and $\theta^{(2)}$ corresponds to the remaining $(d-k)$ components. The profile log likelihood for $\theta^{(1)}$ is now define as
\[\ell_p(\theta^{(1)})=\max_{\theta^{(2)}}\ell(\theta^{(1)},\theta^{(2)}).\]
Then, under suitable regularly conditions, for large n,
\[D_p(\theta^{(1)})=2\{\ell(\hat{\theta}_0)-\ell_p(\theta^{(1)})\}\sim \chi_k^2.\]
For a single component $\theta_i,$ $C_\tau=\{\theta_i:D_p(\theta_i)\leq c_\tau\}$ is a $(1-\tau)$ confidence interval, where $c_\tau$ is the $(1-\tau)$ quantile of the $\chi_1^2$ distribution. If $k=1$ this reduces to the previous definition.

Another method of model selection is the Akaike Information Criterion (AIC). The AIC has played a significant role in solving problems in a wide variety of fields as a model selection criterion for analyzing actual data. The AIC is defined by
\[AIC=-2(\text{maximum log-likelihood})+2(\text{number of free parameters}).\]
The number of free parameters in a model refers to the dimensions of the parameter vector $\theta$ contained in the specified model $f(x|\theta).$ Further, Akaike (1974) states that if the true distribution that generated the data exists near the specified parametric model, the bias associated with the log-likelihood of the model based on the maximum likelihood method can be approximated by the number of parameters.

\section{Inference for GEV distributions.} A potential difficulty with the use of likelihood methods for the GEV concerns the regularity conditions that are required for the usual asymptotic properties associated with the maximum  likelihood estimator to be valid. Such conditions are not satisfied by the GEV model because the end points of the GEV distribution are functions of the parameter values, $\mu-\sigma/\xi$ is an upper end-point of the distribution when $\xi<0,$ and a lower end-point when $\xi>0.$ This violation of the usual regularity conditions means that the standard asymptotic likelihood results are not automatically applicable. Smith (1985) studied this problem in detail and obtained the following results:
\begin{itemize}
\item[(i)] when $\xi>-0.5,$ maximum likelihood estimators are regular, in the sense of having the usual asymptotic properties,
\item[(ii)] if $-1<\xi<-0.5,$ maximum likelihood estimators are generally obtainable, but do not have the standard asymptotic properties, and
\item[(iii)] when $\xi<-1,$ maximum likelihood estimators are unlikely to be obtainable.
\end{itemize}
	Under the assumption that $X_1,\ldots,X_m$ are independent rvs having the GEV distribution, the log likelihood for the GEV parameters when $\xi\neq 0$ is
\begin{eqnarray}\label{ml}
	\ell(x;\mu,\sigma,\xi)&=&-m\log\sigma-(1+1/\xi)\sum_{i=1}^{m}\log\left[1+\xi\left(\dfrac{x_i-\mu}{\sigma}\right)\right]-\sum_{i=1}^{m}\left[1+\xi\left(\dfrac{x_i-\mu}{\sigma}\right)\right]^{-1/\xi},\nonumber\\
&&\text{where, }1+\xi\left(\dfrac{x_i-\mu}{\sigma}\right)>0,\;\;for\;\;i=1,\ldots,m.
\end{eqnarray}
The case $\xi=0$ requires separate treatment using the Gumbel limit of the GEVD. This leads to the log likelihood
\begin{eqnarray}\label{mlg}
	\ell(x;\mu,\sigma)=-m\log\sigma-\sum_{i=1}^{m}\left(\dfrac{x_i-\mu}{\sigma}\right)-\sum_{i=1}^{m}e^{-\left(\frac{x_i-\mu}{\sigma}\right)}.
\end{eqnarray}
There is no analytical solution, but for any given dataset the maximization is straightforward using standard numerical optimization algorithms.

Estimates of extreme quantiles of the maximum distribution under linear normalization are obtained by inverting equation (\ref{gev}):
\begin{eqnarray}\label{As.e2}
x_p=\left\lbrace
\begin{array}{l l}
	\mu-\dfrac{\sigma}{\xi}(1-(-\log(1-p))^{-\xi}),&\;\;\xi\neq 0;\\
	\mu-\sigma \log(-\log(1-p)),&\;\;\xi=0.\\
\end{array}
\right.
\end{eqnarray}
The return levels are exceeded by the annual maximum in any particular time with probability $(1-p).$
If $x_p$ are plotted against $1/(1-p)$ the plots are linear. By substituting the maximum likelihood estimates of the GEV parameters into (\ref{As.e2}), the maximum likelihood estimate of $x_p$ for $0<p<1,$ is obtained as
\begin{equation}\label{IRL.e1}
\hat{x}_p=\left\lbrace
\begin{array}{l l}
	\hat{\mu}-\dfrac{\hat{\sigma}}{\hat{\xi}}(1-y_p^{-\hat{\xi}}),&\;\;\hat{\xi}\neq 0;\\
	\hat{\mu}-\hat{\sigma}\log y_p,&\;\;\hat{\xi}=0,\\
\end{array}
\right.
\end{equation}
where, $y_p=-\log(1-p).$

Furthermore, by the delta method,
\begin{eqnarray}\label{v_MLE}
Var(x_p)\simeq\nabla x_{p}^{T}V_{\theta}\nabla x_{p}.
\end{eqnarray}
where, $\theta=[\mu,\sigma,\xi],$ and $V_{\theta}$ is variance covariance matrix, and 
\begin{eqnarray}
	\nabla x_{p}^{T}&=&\left[\dfrac{\partial x_p}{\partial\mu},\dfrac{\partial x_p}{\partial\sigma},\dfrac{\partial x_p}{\partial\xi}\right],\nonumber\\
	&=&\left[1,-\xi^{-1}(1-y_{p}^{-\xi}),\sigma\xi^{-2}(1-y_{p}^{-\xi})-\sigma\xi^{-1}y_{p}^{-\xi}\log(y_p)\right].\nonumber
\end{eqnarray}
evaluated at $(\hat{\mu},\hat{\sigma},\hat{\xi}).$\\
	If $\hat{\xi}= 0$ and (\ref{v_MLE}) is still valid with
\begin{eqnarray}
\nabla x^{T}_{p}=[1,-\log y_p],
\end{eqnarray}
evaluated at $(\hat{\mu},\hat{\sigma}).$


\textbf{Profile likelihood.} Numerical evaluation of the profile likelihood for any of the individual parameters $\mu,\sigma$ or $\xi$ is straightforward.
For example, to obtain the profile likelihood for $\xi,$ we fix $\xi=\xi_0,$ and maximize the log likelihood (\ref{ml}) with respect to the remaining parameters, $\mu$ and $\sigma.$ This is repeated for a range of values of $\xi_0.$ This methodology can also be applied when inference is required on some combination of parameters. In particular, we can obtain the profile likelihood for any specified return level $x_p.$ This requires a re-parameterization of the GEV model, so that $x_p$ is one of the model parameters, after which the profile log likelihood is obtained by maximization with respect to the remaining parameters in the usual way. Re-parameterization is straightforward,
\begin{eqnarray}\label{e7}
\mu=\left\lbrace
\begin{array}{l l}
 x_p+\dfrac{\sigma}{\xi}[1-y_p^{-\xi}],&\;\;\xi\neq 0;\\
 x_p-\sigma\log y_p,&\;\;\xi= 0,
\end{array}
\right.
\end{eqnarray}

\noindent so that replacement of $\mu$ in (\ref{ml})/(\ref{mlg}) with (\ref{e7}) has the desired effect of expressing the GEV model in terms of the parameters $(x_p,\sigma,\xi).$

\textbf{Model validity.} A probability plot is a comparison of the empirical and fitted distribution functions. With ordered block maximum data $x_{(1)}\leq x_{(2)}\leq\ldots\leq x_{(m)},$ the empirical distribution function evaluated at $x_{(i)}$ is given by
\[\tilde{G}(x_{(i)})=\frac{i}{m+1}.\]
By substitution of parameter estimates into (\ref{gev}), the corresponding model based estimates are
\[\hat{G}(x_{i})=\left\lbrace
 \begin{array}{l l} 
e^{-\left(1+\hat{\xi}\left(\frac{x_{(i)}-\hat{\mu}}{\hat{\sigma}}\right)\right)^{-1/\hat{\xi}}}, & \hat{\xi}\neq 0;\\
e^{-e^{-\left(\frac{x_{(i)}-\hat{\mu}}{\hat{\sigma}}\right)}}, & \hat{\xi}=0.
 \end{array}\right.\]
We then construct plot consisting of the points
\[\Big\{\left(\tilde{G}(x_{(i)}),\hat{G}(x_{(i)})\right),\;\;i=1,\ldots,m\Big\}\]
A weakness of the probability plot for extreme value models is that both $\hat{G}(x_{(i)})$ and $\tilde{G}(x_{(i)})$ are bound to approach 1 as $x_{(i)}$ increases, while it is usually the accuracy of the model for large values of $x$ that is of greatest concern. That is, the probability plot provides the least information in the region of most interest. This deficiency is avoided by the quantile plot, consisting of the points
\[\Big\{\left(\hat{G}^{-1}(i/(m+1)),x_{(i)}\right),\;\;i=1,\ldots,m\Big\}.\]
If $\hat{G}$ is a reasonable estimate of $G,$ then the quantile plot should also consist of points close to the unit diagonal.

\section{Results}
\begin{figure}[!ht]
	\caption{Profile likelihood vs Shape Parameter.}
	\label{profile_shape}
	\includegraphics[width=8cm]{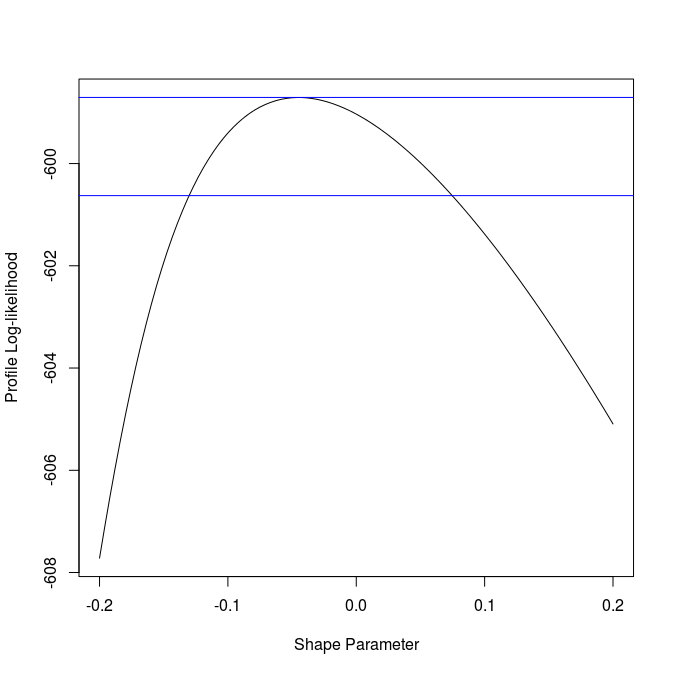}
\end{figure}

This analysis is based on the annual maximum yearly rainfall data of station Eudunda, Australia which collected during 1881-2013. Here, we analyse the data for 1881-2009 and we control the model with real values of maximum rainfall as $[106.2,\,104,\,60.8,\,73.8]$ correspond to the years $2010-2013.$ It seems reasonable to assume that the pattern of variation has stayed constant over the observation period, so we model the data as independent observations from the GEV distribution.
Maximization of the GEV log likelihood using the "Nelder-Mead (1965)" we get
\[(\hat{\mu},\hat{\sigma},\hat{\xi})=(79.24932448, 22.11954846, -0.04483979),\]
with standard error $2.1507407, 1.5202669$ and $0.0519857$ respectively.
The negative log-likelihood is $598.7072,$ and Fig.\ref{profile_shape} shows the profile log-likelihood for $\xi.$

\begin{figure}[!ht]
\center
	\caption{Diagnostic plots for Gumbel fit to the data}
		\label{modelGumbel}
	\includegraphics[width=15cm]{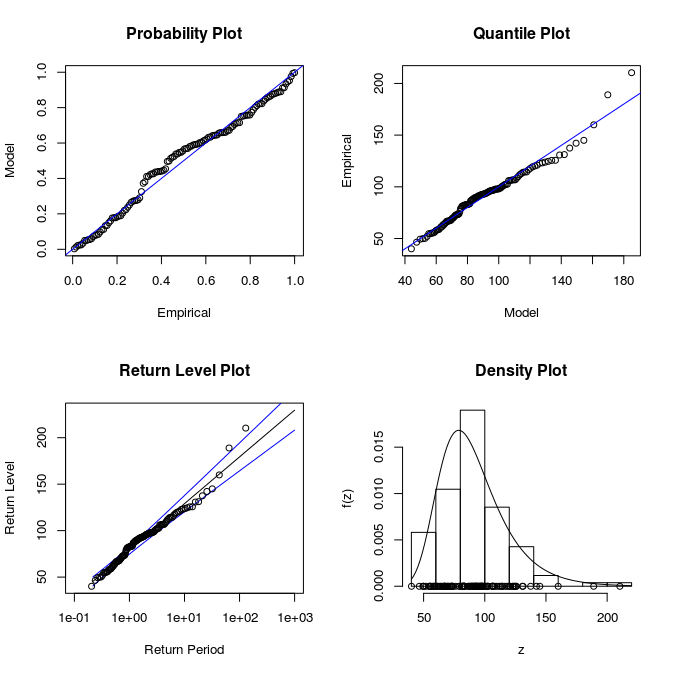}
\end{figure}

Maximum likelihood in the Gumbel case corresponds to maximization of (\ref{mlg}) giving

\[(\hat{\mu},\hat{\sigma})=(78.70124, 21.85684),\]
with standard error $2.031079$ and $1.471843$ respectively. The negative of the maximized log-likelihood is $599.0322.$ The likelihood ratio test statistic for the model between the Gumbel and extreme value with $\hat{\xi}=-0.04483979$ is,
\[D=2(599.0322-598.7072)=0.6498759,\]
 These values are small when compared to the $\chi^2_1=3.84$ suggesting  that the Gumbel model is adequate for these data.
Also, from Table.\ref{ML_xi}, the value of  AIC in the Gumbel model is lower than the value of AIC for GEV when $\xi$ is estimated by maximum likelihood method. So, the Gumbel model is better than the GEVD.
The various diagnostic plots for assessing the accuracy of the Gumbel model fitted to the data are shown in Fig.\ref{modelGumbel}. 
 The probability plot and the quantile plot give cause to doubt the validity of the fitted model, each set of plotted points is near linear. 
The return level curve asymptotes to a finite level as a consequence of the negative estimate of $\xi$ and the estimate of $\xi$ is close to zero, the estimated curve is close to linear. Finally, the corresponding density estimate seems consistent with the histogram of the data. Consequently, all four diagnostic plots lend support to the fitted Gumbel model.

The bootstrap bias and Jackknife bias calculate in Table.\ref{bias2}. Since, the ratio of Jackknife bias to standard error is greater than $0.25$ for parameter $\sigma$ then, the correction value of parameter is obtained by
\[\hat{\sigma}_{\text{corr}}=\hat{\sigma}-\text{Bias}_{\text{jack}}=21.11317.\]
 The Table.\ref{CI_Return} shows the estimate return level for some years with $95\%$ CIs. The corresponding estimate for the $4$ years return level is $\hat{x}_{0.25}=105.0061,$ with a $95\%$ CI $[99.83843,110.1738]$ and the real values of maximum rain fall for the period $2010-2013$ is $106.2,$ which has good accuracy in the real value and predict value at 4 years.

\textbf{Further discussion.}
Suppose that the maximum yearly rainfall $X,$ has cdf
\[F(x)=e^{-e^{-\frac{x-\hat{\mu}}{\hat{\sigma}_{corr}}}},\;\;x\in\Real.\]
The pdf of the $rth$ order statistic in a sample of size $n$ is
	\begin{eqnarray}
		f_{X_{r:n}}(x)&=&r\binom{n}{r}f(x)[F(x)]^{r-1}[1-F(x)]^{n-r},\nonumber
	\end{eqnarray}
and the corresponding cdf is
\begin{eqnarray}
	F_{X_{r:n}}(x)&=&\sum_{k=r}^{n}\binom{n}{k}\left[F(x)\right]^{k}\left[1-F(x)\right]^{n-k}.\nonumber
\end{eqnarray}
Suppose that a value of rainfall is $x=100mm.$ The probability of $5th$ order statistics for the value of the rainfall to be less than $100$ in a period of 10 years is
\[F_{X_{5:10}}(100)=0.94843246.\]
Table. \ref{order} shows the different probability for different order statistics in a period of 10 years.

\begin{table}[!ht]
		\centering
\caption{Diagnostic model for different values of $\xi$}\label{ML_xi}
		\begin{tabular}{p{2.5cm} p{4cm} p{2cm}}
			\hline
			  $\hat{\xi}$ & MLE  & AIC  \\
			\hline
			$0$ & $-599.0322$ & $1202.064$\\
			$-0.04483979$ & $-598.7072$ &$1203.414$\\
			\hline
		\end{tabular}
		\end{table}

\begin{table}[!ht]
	\centering
\caption{The Bias and standard error values for parameters of Gumbel model}\label{bias2}
    \begin{tabular}{  p{1.5cm} p{3.5cm} p{2cm}p{3cm} p{3cm} p{2cm} }
    \hline
    Parameter &  Bias(SE) & Bias/SE &$\text{Bias}_{\text{jak}}(\text{SE}_{\text{jack}})$&$\text{Bias}_{\text{jak}}/\text{SE}_{\text{jak}}$ \\ \hline
    $\mu$  & $0.05594263(2.05307)$ & $0.02724828$ & $0.1488887 (7.171212)$ &$0.02076199$ \\
    $\sigma$ &$-0.07724028(1.337519)$ & $0.05774891$ & $0.7436685(2.341009)$ & $0.31767007$\\
    \hline
    \end{tabular}
\end{table}
		
\begin{table}[!ht]
		\centering
		\caption{Return level for Gumbel Distribution}\label{CI_Return}
		\begin{tabular}{p{3cm} p{3cm} p{3cm} p{2.5cm}}
			\hline
			  Return Period & Return Level & Lower bound & Upper bound \\
			\hline
$4$ & $105.0061$ & $ 99.83843$ & $110.1738$ \\
$10$ & $126.2136$ & $118.96894$ & $133.4583$\\
$40$ & $156.3185$ & $145.85938$ & $166.7775$\\
$100$ & $ 175.8250$ & $163.21585$ & $ 188.4341$\\
			\hline
		\end{tabular}
			\end{table}
		
	\begin{table}[!ht]
		\centering
		\caption{$rth$ order statistic for the rainfall value $100mm$}		
\label{order}
		\begin{tabular}{p{3cm} p{3cm}}
			\hline
			  $r$ & $P(X_{r:10}<100)$\\
			\hline
			$2$ & $0.99983162$ \\ 
			$4$&  $0.98818640$\\
			$5$ & $0.94843246$\\
			$8$&  $0.36806786$\\
			$10$& $0.02607971$\\
			\hline
		\end{tabular}
\end{table}	

\section*{aknowlegment}
Thanks to the Government of Australia, Bureu of Meteorology, who made the datasets used here freely available on their website.

\section{The R program}
\begin{verbatim}
library(nlme); library(mgcv)
library(bootstrap) ; library(ismev)
xm<-read.table("~//S19.txt",header=TRUE)
S1<-as.numeric(xm$data) ; Year<-xm$Year; m<-length(S1)
#--------------GEV-------------------
ge<-gev.fit(S1)
ge_mu<-ge$mle[1]; ge_sigma<-ge$mle[2]; ge_xi<-ge$mle[3]
gev.diag(ge)
gev.profxi(ge,-0.2,0.2,conf=0.95,nint=100)
savePlot("profile_shape.jpg")
#--------------Gumbel-------------------
gu<-gum.fit(S1)
gu_mu<-gu$mle[1]; gu_sigma<-gu$mle[2]
gum.diag(gu)
savePlot("modelGumbel.jpg")
#--------- Check the models----------
Ratio<-2*(gu$nllh-ge$nllh)
AIC1<--2*(-gu$nllh)+2*2 ; AIC2<--2*(-ge$nllh)+2*3
list(Ratio=Ratio, AIC_gumbel=AIC1, AIC_Gev=AIC2)
#------------ Bootstrap for Gum------------
theta2<-function(S){gum.fit(S)$mle}
R<-999
gu_boot<-bootstrap(S1,R,theta2)
gu_mub<-mean(gu_boot$thetastar[1,])-gu_mu
gu_sigmab<-mean(gu_boot$thetastar[2,])-gu_sigma
list(bias.mu=gu_mub, bias.sigma=gu_sigmab)
list(ratio.mu=abs(gu_mub/sd(gu_boot$thetastar[1,]))
      ,ratio.sigma=abs(gu_sigmab/sd(gu_boot$thetastar[2,])))
#---------- Jackknife Method for Gum-------------------
u2<-array(0,c((m-1),2)); jack.bias<-numeric(0)
jack.se<-numeric(0)
for(i in 1:(m-1)){
	for(j in 1:2)
        u2[i,j] <- gum.fit(S1[-(i+1)])$mle[j]}
	for(k in 1:2) {
	    jack.bias[k] <- (m - 1) * (mean(u2[,k]) - gu$mle[k])
        uu<-sum(u2[,k])/m
    for(k in 1:2)
	    jack.se[k] <- sqrt(((m - 1)/m) * sum((u2[,k] - uu)^2))}
list(jack.se = jack.se, jack.bias = jack.bias, ratio=jack.bias/jack.se)
 gu_sigmaC<-gu_sigma-jack.bias[2]
#-------------------Return Value ---------
up_b<-array(0); low_b<-array(0); xp_g<-array(0)
var2<-function(yp){
zp<-c(1,-log(yp))
var_yp<-t(zp)%*%gu$cov%*%zp
var_yp}
i<-1; ret<-c(.25,.1,.025,.01)
 for(p in ret){
      yp<--log(1-p)
      xp_g[i]<-(gu_mu)-(gu_sigmaC)*log(yp)
      up_b[i]<-xp_g[i]+qnorm(.95)*sqrt(var2(yp))
      low_b[i]<-xp_g[i]-qnorm(.95)*sqrt(var2(yp))
      i<-i+1}
list(return_period=1/ret, return_level=xp_g)
list(Lower_bound=low_b, Upper_bound=up_b)
#---------------Order Statistics-----------
s2<-array(0); F<-exp(-exp(-(100-gu_mu)/gu_sigmaC))
n<-10;i<-1
  for(r in c(2,4,5,8,10)){
s<-0
  for(k in r:n)
s<-s+(choose(n,k)*F^k*(1-F)^(n-k))
s2[i]<-s; i<-i+1}
list(Prob_order=s2)
\end{verbatim}

\end{document}